\documentclass[aps,pra,twocolumn,preprintnumbers,amsmath,amssymb,showpacs]{revtex4}
\usepackage{graphicx,epsfig,epsf}
\usepackage{dcolumn}
\usepackage{bm}
\usepackage{mathbbol}

\begin{document}
\newcommand{\ri}{{\rm i}}
\newcommand{\re}{{\rm e}}
\newcommand{\bx}{{\bf x}}
\newcommand{\bd}{{\bf d}}
\newcommand{\br}{{\bf r}}
\newcommand{\bk}{{\bf k}}
\newcommand{\bE}{{\bf E}}
\newcommand{\bR}{{\bf R}}
\newcommand{\bM}{{\bf M}}
\newcommand{\bn}{{\bf n}}
\newcommand{\bs}{{\bf s}}
\newcommand{\tbs}{\tilde{\bf s}}
\newcommand{\rSi}{{\rm Si}}
\newcommand{\beps}{\mbox{\boldmath{$\epsilon$}}}
\newcommand{\rg}{{\rm g}}
\newcommand{\tr}{{\rm tr}}
\newcommand{\xmax}{x_{\rm max}}
\newcommand{\ra}{{\rm a}}
\newcommand{\rx}{{\rm x}}
\newcommand{\rs}{{\rm s}}
\newcommand{\rP}{{\rm P}}
\newcommand{\up}{\uparrow}
\newcommand{\down}{\downarrow}
\newcommand{\hc}{H_{\rm cond}}
\newcommand{\kb}{k_{\rm B}}
\newcommand{\cI}{{\cal I}}
\newcommand{\tit}{\tilde{t}}
\newcommand{\cE}{{\cal E}}
\newcommand{\cC}{{\cal C}}
\newcommand{\Ubs}{U_{\rm BS}}
\newcommand{\qq}{{\bf ???}}
\newcommand*{\etal}{\textit{et al.}}
\sloppy

\title{Coherent control of atomic tunneling}
\author{John Martin and Daniel Braun}
\affiliation{Laboratoire de Physique Th\'eorique, IRSAMC, UMR 5152 du CNRS,
  Universit\'e Paul Sabatier, Toulouse, FRANCE}

\begin{abstract}
We study the tunneling of a two-level atom in a double well
potential while the atom is coupled to a single electromagnetic
field mode of a cavity. The coupling between internal and external
degrees of freedom, due to the mechanical effect on the atom from
photon emission into the cavity mode, can dramatically change the
tunneling behavior. We predict that in general the tunneling process
becomes quasiperiodic. In a certain regime of parameters a collapse
and revival of the tunneling occurs. Accessing the internal degrees
of freedom of the atom with a laser allows to coherently manipulate
the atom position, and in particular to prepare the atom in one of
the two wells. The effects described should be observable with atoms
in an optical double well trap.
\end{abstract}

\maketitle

\section{Introduction}
The tunneling effect is considered one of the hallmarks of quantum
mechanical behavior. Historically, tunneling was first examined for
single particles (e.g.\ $\alpha$ particles \cite{Gamow28}, electrons
in field emission \cite{Guth41} and later in mesoscopic circuits
\cite{Devoret92}), for Cooper pairs \cite{Josephson74}, and for
molecular groups \cite{Hueller80,Wuerger89,DBraun94}. Recently the
tunneling of atoms has attracted substantial attention
\cite{Louis95,Meier01,Luxat02,Albiez05}. Dynamical (chaos assisted)
tunneling of ultracold atoms between different islands of stability
in phase space was analyzed in \cite{Grossmann91,Averbukh02} and has
been observed experimentally~\cite{Steck01,Hensinger01}. Resonantly
enhanced tunneling of atoms between wells of a tilted optical
lattice has also been observed very recently \cite{Sias07}. In all
of these examples, the atoms have been considered internally as
inert, and only the center of mass coordinate of the atom was of
interest. In \cite{Haycock00} it was shown that by taking into account the
internal degrees of freedom of atoms, an atom/optical double well
potential could be created in which tunneling atoms see their
internal and external states correlated (such an effect is also known
from other contexts \cite{Salzburger04}). Mechanical
effects of  
light in optical resonators were also investigated in
\cite{Domokos03}, but no tunneling was considered. 

Here we show that
the tunneling effect can be drastically modified if an internal
transition of the atom is coupled to a single electromagnetic mode
in a cavity, such that photon emission is a reversible and coherent
process. The resulting Rabi oscillations between states with the
excitation in the atom and states with a photon in the cavity
modulate the periodic tunneling motion. Depending on the frequencies
involved, a rich quasi-periodic behavior can result. If the cavity
is fed with a coherent state, collapse and revival of the tunneling
effect can occur. Moreover, we show that one may profit from access
to the internal degrees of freedom of the atom (e.g.\ with a laser)
to control the atomic motion in the external potential.
\begin{figure}
\begin{center}
\includegraphics[width=.95\linewidth]{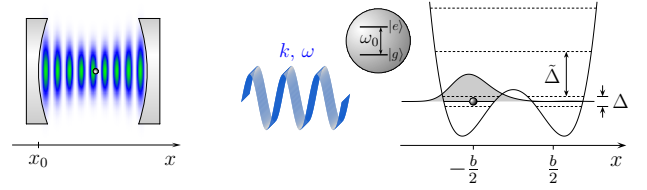}
\end{center}
\caption{(Color online) Two-level atom in a double well potential
interacting with a standing wave inside a cavity.}
\label{doublewell}
\end{figure}

\section{Model}
\subsection{Derivation of the Hamiltonian}
Consider a trapped two-level atom (with levels $|g\rangle$,
$|e\rangle$ of energy $\mp \hbar \omega_0/2$ respectively)
interacting with a standing wave (with wave number $k$ and frequency
$\omega$) inside a cavity as illustrated in Fig.~\ref{doublewell}.
The atom is assumed to be bound in the $y-z$ plane at the
equilibrium position $y=z=0$ and to experience a symmetric double
well potential $V(x)$ along the $x$ direction. We denote by $\Delta$
the tunnel splitting, i.e.\ the energy spacing between the two
lowest energy states (the symmetric $|-\rangle$ and antisymmetric
$|+\rangle$ states) of this double well potential.
 Below we also allow the trapped atom to interact resonantly with
an external laser. The Hamiltonian of this system is given by
\begin{equation}\label{tHamiltonian}
    H= H_{A}+H_{F}+H_{AF},
\end{equation}
where $H_{A}=H_{A}^{\mathrm{ex}}+H_{A}^{\mathrm{in}}$ is the
Hamiltonian of the trapped atom, $H_{F}$ is the Hamiltonian of the
free field and $H_{AF}$ is the interaction Hamiltonian describing
the atom-field interaction. We have
\begin{equation}
\begin{aligned}
    & H_{A}^{\mathrm{ex}}= \frac{p_x^2}{2m}+V(x),\\
    & H_{A}^{\mathrm{in}}=
    \frac{\hbar\omega_0}{2}\sigma_z^{\mathrm{in}},\\
    & H_{F}= \hbar\omega
    a^{\dagger}a,\\
    & H_{AF}= -\mathbf{d.E},
\end{aligned}
\end{equation}
where $\mathbf{d}$ denotes the atomic dipole,
\begin{equation}
    \mathbf{E}=
    E_{\omega}\boldsymbol{\varepsilon}\left(
   a+a^{\dagger}\right)\sin(k(x-x_0))
\end{equation}
is the electric field operator, with $E_{\omega}=\sqrt{\frac{\hbar
\omega}{\epsilon_0V}}$, where $\epsilon_0$ is the permittivity of
free space, $V$ the electromagnetic mode volume, $x_0$ the abscissa
at the left cavity mirror ($x_0<0$), and $\boldsymbol{\varepsilon}$
the electric field polarization vector. We have introduced the
operators $\sigma_i^{\mathrm{in}}$ (resp.\ $\sigma_i^{\mathrm{ex}}$)
for $i=x,y,z$ as the Pauli spin operators in the basis
$\{|e\rangle,|g\rangle\}$ (resp.\ $\{|+\rangle,|-\rangle\}$). The
operator $x$ stands for the center-of-mass position of the atom,
$p_x$ is the conjugate momentum along the $x$ axis, $m$ denotes the
atomic mass, and $a$ ($a^{\dagger}$) the annihilation (creation)
operator of the cavity radiation field.

We adopt the two-level approximation which consists of taking into
account only the two lowest motional energy states. This requires
the Rabi frequency $\sqrt{4g^2+\delta^2}$ (with
$\delta=\omega-\omega_0$ the detuning between the cavity field and
the atomic transition frequencies) to be much smaller than the
frequency gap $\tilde{\Delta}$ between the upper motional states and
the ground state doublet (see Fig.~\ref{doublewell}). Within this
approximation, Hamiltonian $H_{A}^{\mathrm{ex}}$ becomes
\begin{equation}
H_{A}^{\mathrm{ex}}= \frac{\hbar\Delta}{2}\sigma_z^{\mathrm{ex}}
\end{equation}
and the position operator takes the form
$x=\frac{b}{2}\sigma_x^{\mathrm{ex}}$ with $b/2=\langle
+|x|-\rangle$. We can form states that are mainly concentrated in
the left/right wells,
\begin{equation}
\begin{aligned}
|L\rangle ={} (|+\rangle-|-\rangle)/\sqrt{2},\\
|R\rangle ={} (|+\rangle+|-\rangle)/\sqrt{2}.
\end{aligned}
\end{equation}
The average position of a particle localized in the right well is
then given by $b/2$ (see Fig.~\ref{doublewell}) and
$\sigma_x^{\mathrm{ex}}=|R\rangle\langle R|-|L\rangle\langle L|$.
The interaction Hamiltonian $H_{AF}$ can then be written
\begin{equation}\label{HAF2}
    H_{AF}= -\hbar g (a+a^{\dagger})\Big[ \sin\chi\cos\kappa\; \sigma_x^{\mathrm{in}}
     - \cos\chi\sin\kappa\; \sigma_x^{\mathrm{ex}}
     \sigma_x^{\mathrm{in}}\Big]\nonumber
\end{equation}
with the atom-field coupling strength $g=-\langle
e|\mathbf{d}|g\rangle\mathbf{.}\boldsymbol{\varepsilon}
E_{\omega}/\hbar$, and
\begin{equation}
\chi=kx_0, \;\;\; \kappa=kb/2.
\end{equation}
For long wavelengths ($\kappa\ll 1$), or $\kappa=n\pi$ with integer
$n$, the left and right sites of the double well are
indistinguishable to the cavity photon and $H_{AF}$ reduces to
Jaynes-Cummings Hamiltonian without rotating wave approximation
(with a sine varying coupling constant), $-\hbar g\sin\chi\;
(a+a^{\dagger}) \sigma_x^{\mathrm{in}}$. Note that $\kappa\ll 1$
would normally be identified with the Lamb-Dicke regime. Here the
situation is more subtle as the level spacings between the tunneling
split ground state doublet and the next excited states can be very
different such that the recoil energy $\hbar\omega_{\rm recoil}$ satisfies
$\Delta\ll\omega_{\rm recoil}\ll\tilde{\Delta}$. One may thus be in the
Lamb-Dicke regime concerning 
transitions to higher vibrational states but have a significant
mechanical effect on the atomic tunneling. Furthermore, since there
is only one photon mode, the recoil energy cannot vary continuously and
exciting higher vibrational levels requires $\omega_{\rm recoil}$ close to a
level spacing. Our numerical calculations show that even for $\kappa\sim 1$ the
two-level approximation can still work very well (see
Fig.~\ref{rhoLLrhoee}).

For $\delta$, $\Delta \ll \omega$, $\omega_0$, a rotating wave
approximation is justified, which consists in eliminating the energy
non-conserving terms $a \sigma_{\pm}^{\mathrm{ex}}
\sigma_-^{\mathrm{in}}$ and $a^{\dagger} \sigma_{\pm}^{\mathrm{ex}}
\sigma_+^{\mathrm{in}}$ with $\sigma_+^{\mathrm{in}} = |e \rangle
\langle g|$,
$\sigma_-^{\mathrm{in}}=\sigma_+^{\mathrm{in}\,\dagger}$ and
$\sigma_+^{\mathrm{ex}} = |+ \rangle \langle -|$,
$\sigma_-^{\mathrm{ex}}=\sigma_+^{\mathrm{ex}\,\dagger}$. Within
this approximation, the total Hamiltonian reads
\begin{eqnarray}
    H& =&\frac{\hbar\Delta}{2}\sigma_z^{\mathrm{ex}}+\frac{\hbar\omega_0}{2}\sigma_z^{\mathrm{in}}+ \hbar\omega a^{\dagger}a\label{tHamiltonianf}\\
&&+\hbar g
    (a\sigma_+^{\mathrm{in}}+a^{\dagger}\sigma_-^{\mathrm{in}})\Big[\cos\chi\sin\kappa\;\sigma_x^{\mathrm{ex}}-  \sin\chi\cos\kappa\; \mathbb{1}^{\mathrm{ex}}\Big]\,.\nonumber
\end{eqnarray}
 Thus, depending on
the parameters $\chi$ and $\kappa$, the cavity photon may induce
internal transitions in the atom only ($\cos\chi\sin \kappa=0$), or
induce transitions between internal and external states at the same
time ($\cos\chi\sin \kappa\ne 0$) even for a vanishing detuning
($\delta=\omega-\omega_0=0$). This is in contrast to conventional
sideband transitions of harmonically bound atoms or ions in the
Lamb-Dicke regime which require an appropriate value of the
detuning. For a fixed potential center (and thus fixed $\chi$),
$\kappa$ can be changed through a modulation of the well-to-well
separation $b$. We will neglect in the following the effects of
decoherence, which means that not only $g$ but also $\Delta$ should
be much larger than the rate of spontaneous emission $\Gamma$, and
the cavity decay rate $\kappa_{\rm cav}$.

We denote  the global state of the atom-field system by
$|n,i,j\rangle\equiv |n\rangle\otimes|i\rangle\otimes|j\rangle$
where $|n\rangle$ stands for the cavity field eigenstates,
$|i\rangle\in\{|-\rangle,|+\rangle\}$ for the external motional
states, and $|j\rangle\in\{|g\rangle,|e\rangle\}$ for the internal
states. The total excitation number $N$ is given by
$a^{\dagger}a+\sigma_+^{\mathrm{in}}\sigma_-^{\mathrm{in}}$.

\subsection{Energy levels} The states
$|0,\pm,g\rangle$ are eigenstates of $H$ with eigenvalue
$(-\hbar\omega_0\pm \hbar\Delta)/2$, i.e.~these states remain
uncoupled and represent the two lowest energy states in the regime
$\delta$, $\Delta \ll \omega$, $\omega_0$. It is straightforward to
verify that the Hamiltonian (\ref{tHamiltonianf}) only induces
transitions between states with the same number of excitations $N$,
$\{ |N-1,+,e\rangle, |N,+,g\rangle, |N-1,-,e\rangle, |N,-,g\rangle
\}\equiv \{ |1\rangle, |2\rangle, |3\rangle, |4\rangle\}$. It is
therefore sufficient to solve the dynamics in this subspace. In
doing so, we obtain the eigenvalues of $H$,
\begin{equation}
\lambda_{\rho\mu}=(N-1/2)\hbar\omega+\rho\frac{\hbar \Omega_\mu}{2},
\end{equation}
for $\rho,\mu\in \{\pm\}$, $N=1,2,\hdots$, and with
\begin{widetext}
\begin{eqnarray}
    \label{Opm} & \Omega_{\pm} &=\sqrt{2Ng^2(1-\cos(2\kappa)\cos(2\chi))+\delta^2+\Delta^2\pm 2\Omega^2}\,,\\
    \label{O2} & \Omega^2 &=\sqrt{4Ng^2\cos^2\kappa\sin^2\chi(\Delta^2+4Ng^2\sin^2\kappa\cos^2\chi)+\delta^2\Delta^2}\,.
\end{eqnarray}
\end{widetext}
For a vanishing tunnel splitting ($\Delta= 0$),
$\Omega_{\pm}$ reduces to the maximum (minimum) of the two Rabi
frequencies of the Jaynes-Cummings models in the right and left
wells. For $\cos\kappa=1$, the decoupling of external and internal
degrees of freedom manifests itself also in the eigenvalues with
$\Omega_{\pm}=|\sqrt{4Ng^2\sin^2\chi+\delta^2}\pm \Delta|$.

\subsection{Evolution operator}
The whole dynamics of the system can be described by means of the
evolution operator $U(t)= e^{-iHt/\hbar}$ with components
$U_{ij}=\langle i|U(t)|j\rangle=U_{ji}$, which can be calculated
exactly. In order to simplify the expressions, we restrict ourselves
in the following to $\chi=-\pi/4-2n\pi$ (integer $n$). We find, up
to a an overall phase $e^{-i(N-1/2)\omega t}$,
\begin{equation}
\begin{aligned}
    U_{11} = -\frac{i}{2
    \Lambda}\sum_{\mu=\pm}\Big[&\mu S_\mu\Omega_{-\mu}\left\{\xi+\mu(\Delta-\delta)\Omega^2\right\}\\
    &-i\mu\Omega_+\Omega_-C_\mu(\delta\Delta-\mu\Omega^2)\Big],
\end{aligned}
\end{equation}
\begin{equation}
\begin{aligned}
    U_{12} = \frac{-i\sqrt{N}g\cos\kappa}{\sqrt{2} \Lambda}
    \sum_{\mu=\pm}\Big[&\mu S_\mu\Omega_{-\mu}(\Delta^2+2Ng^2\sin^2\kappa\\&+\mu\Omega^2)
    +i\mu\Omega_+\Omega_-\Delta C_\mu\Big],
\end{aligned}
\end{equation}
\begin{equation}
\begin{aligned}
    U_{13} ={}& \frac{-iNg^2\sin (2\kappa)}{2 \Lambda}
    \sum_{\mu=\pm}\big[\mu\delta\Omega_\mu
    S_{-\mu}+i\mu\Omega_+\Omega_-C_\mu
    \big],
\end{aligned}
\end{equation}
\begin{equation}
\begin{aligned}
    U_{23} ={}& \frac{i\sqrt{N}g \sin\kappa}{\sqrt{2} \Lambda}
    \sum_{\mu=\pm}\big[\mu \Omega_\mu S_{-\mu}(\delta\Delta+2Ng^2\cos^2\kappa-\mu\Omega^2)\big]
\end{aligned}
\end{equation}
with
\begin{equation}
\begin{aligned}
    &\xi=\Delta(\delta^2+2Ng^2\cos^2\kappa-\delta\Delta),\\
    & \Lambda=\Omega_{+} \Omega_{-} \Omega^2,
\end{aligned}
\end{equation}
and where all time dependence is in the coefficients
\begin{equation}
    C_{\pm}=\cos(\Omega_{\pm} t/2),\;\;\; S_{\pm}=\sin(\Omega_{\pm} t/2).
\end{equation}
The remaining components can be deduced from the relations
$U_{22}(\delta,\Delta)=U_{33}(-\delta,-\Delta)=U_{44}(\delta,-\Delta)=U_{11}(-\delta,\Delta)$,
$ U_{24}(\delta,\Delta)= U_{13}(-\delta,\Delta)$, $
U_{14}(\delta,\Delta)= U_{23}(\delta,-\Delta) =
U_{23}(-\delta,\Delta)$, and $   U_{34}(\delta,\Delta)=
U_{12}(\delta,-\Delta)$, valid for any $\chi$, where we have made
explicit the dependence of the $U_{ij}$ on $\delta$ and $\Delta$.

\section{Internal and external dynamics}

The reduced density matrix $\rho^{\mathrm{ex}}$ for the atomic
center-of-mass motion alone follows from
$\rho=|\psi(t)\rangle\langle\psi(t)|$ by tracing out the field and
internal degrees of freedom, where the total wave function at time
$t$ reads $|\psi(t)\rangle = \sum_{i,j=1}^{4}U_{ij}\langle
    j|\psi(0)\rangle\:|i\rangle$. The average position of
the atom in the double well potential is then given by
\begin{equation}
\langle x
    \rangle=\frac{b}{2}\,
    \mathrm{Tr}_{\mathrm{ex}}(\rho^{\mathrm{ex}}\sigma_x^{\mathrm{ex}})=
    \frac{b}{2}(1-2\rho_{LL})
\end{equation}
with $\rho_{LL}=\langle L|\rho^{\mathrm{ex}}|L\rangle$. Similarly,
we obtain the reduced density matrix $\rho^{\mathrm{in}}$ for the
internal atomic state by tracing out the field and external degrees
of freedom, and the probability to find the atom in the excited
state as $\rho_{ee}=\langle e|\rho^{\mathrm{in}}|e\rangle$.

In the following, we first focus on resonant atom-field interaction
($\omega=\omega_0$) before moving to the non-resonant case
($\omega\ne \omega_0$). We distinguish three regimes according to
the tunnel splitting compared to the Rabi frequency $g$~: the small
tunnel splitting regime (when $\Delta/g\ll 1$), the intermediate
regime (when $\Delta/g\sim 1$), and the large tunnel splitting
regime (when $\Delta/g\gg 1$).

\subsection{Resonant atom-field interaction}

For resonant atom-field interaction ($\delta=0$), the expressions
for $U_{ij}$ can be greatly simplified. If the system is initially
prepared in the state $|N-1,R,e\rangle$ and for $\kappa=\pi/4$, we
have
\begin{equation}\label{rhoLLresonant}
    \rho_{LL}=\frac{\Delta^2}{\Delta^2+Ng^2}\sin^2\left(\frac{\Omega_{\mathrm{tun}}
    t}{2}\right)
\end{equation}
with the tunnel frequency
\begin{equation}\label{tunfreq}
    \Omega_{\mathrm{tun}}=
    \frac{1}{2}\left(\Omega_++\Omega_-\right),
\end{equation}
and
\begin{equation}\label{rhoeeresonant}
    \rho_{ee}=\frac{1}{2}+
    \frac{\sum_{\mu=\pm}\left(\Omega_\mu^2-\Delta^2\right)\cos(\Omega_\mu t)
    +4\Delta^2\cos\left(\frac{\Omega_+-\Omega_-}{2}t\right)}{8(Ng^2+\Delta^2)
    }.
\end{equation}

The atom position oscillates with a single frequency $
\Omega_{\mathrm{tun}}$ given by Eq.~(\ref{tunfreq}), whereas
$\rho_{ee}$ evolves with three in general incommensurable
frequencies $\Omega_+$, $\Omega_-$, and $(\Omega_+-\Omega_-)/2$
giving rise to a quasi-periodic signal.

For $\Delta/g\ll 1$, Eq.~\eqref{rhoLLresonant} leads to
$\rho_{LL}\simeq 0$ (up to order $(\Delta/g)^2$), indicating that
tunneling is suppressed. This is already obvious from
(\ref{tHamiltonianf}), as the term responsible for tunneling,
$(\hbar\Delta/2)\sigma_z^{\mathrm{ex}}=(\hbar\Delta/2)(|R\rangle\langle
L|+|L\rangle\langle R|)$ becomes very small compared to the last
term, diagonal in $|R\rangle,|L\rangle$ which leads to internal Rabi
flopping. Note, however, that tunneling is suppressed on all time
scales, even for $t\gg 1/\Delta$,  due to the reduced amplitude in
Eq.~(\ref{rhoLLresonant}), very much in contrast to tunneling
without internal degrees of freedom, where only the period of the
tunneling motion, but not the amplitude is affected when $\Delta$ is
reduced. For $\kappa$ approaching $\pi$, the situation changes
because the term $g\cos\chi\sin\kappa\;\sigma_x^{\mathrm{ex}}$ of
the interaction Hamiltonian inducing transitions between vibrational
states becomes small in comparison with $\Delta$ thereby allowing
tunneling again.

\begin{figure}
\begin{center}
\includegraphics[width=.8\linewidth]{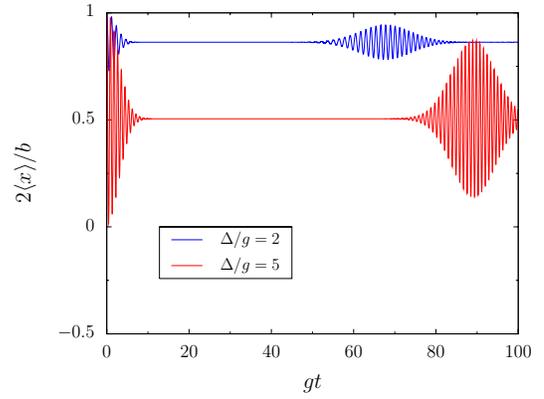}
\end{center}
\caption{(Color online) Average position of the atom in the double
well as a function of time for $\Delta/g=2$ (blue, top curve) and
$\Delta/g=5$ (red), $\kappa=\pi/4$ and a coherent state with
$\alpha=5$.} \label{rhoLL_collapse}
\end{figure}

Because internal and external degrees of freedom are coupled, the
tunneling frequency (Eq.~(\ref{tunfreq})) depends on the number of
photons inside the cavity. As an example, let us now consider
$\Delta\sim g$ and a cavity field initially in a coherent state $
|\alpha\rangle=e^{-\frac{1}{2}|\alpha|^2}\sum_{n=0}^{\infty}\frac{\alpha^n}{\sqrt{n!}}|n\rangle$
with $|\alpha|^2$ equal to the mean photon number $\langle n
\rangle$. Figure~\ref{rhoLL_collapse} shows that the average
position of the atom in the double well as a function of time for a
coherent state exhibits collapses and revivals. The oscillation
amplitude decreases with increasing mean photon number $\langle n
\rangle$ and decreasing tunnel splitting $\Delta$ (see
Eqs.~(\ref{rhoLLresonant},\ref{Opm})). Since the probability to find
the atom in the excited state oscillates with three frequencies, no
collapses and revivals are observed for $\rho_{ee}$.

The collapse time $t_c$ of the tunneling motion can be estimated
from the condition~\cite{Scully97} $(\Omega_{\mathrm{tun}}(\langle n
\rangle+\sqrt{\langle n \rangle})-\Omega_{\mathrm{tun}}(\langle n
\rangle-\sqrt{\langle n \rangle}))\,t_c\sim 1$ with
$\Omega_{\mathrm{tun}}(m)$ given by Eq.~\eqref{tunfreq} for $N=m+1$,
which yields, for $\langle n \rangle\gg 1$,
\begin{equation}\label{tc}
    t_c\sim
    \frac{1}{g}\left(1+\frac{(\Delta/g)^2+3/4}{2\langle n \rangle}\right)+\mathcal{O}(\langle n \rangle^{-2})
\end{equation}

The time interval between two
following revivals, $t_r$, follows from
$\left(\Omega_{\mathrm{tun}}(\langle n \rangle)-\Omega_{\mathrm{tun}}(\langle n \rangle-1)\right)\,t_r
= 2\pi$, and is given for $\langle n \rangle\gg 1$ by
\begin{equation}\label{tr}
    t_r\simeq
    \frac{4\pi\sqrt{\langle n \rangle}}{g}\left(1+\frac{(\Delta/g)^2+1/2}{2\langle n \rangle}+\mathcal{O}(\langle n \rangle^{-2})\right)
\end{equation}
For the parameters of Fig.~\ref{rhoLL_collapse}, Eq.~\eqref{tr}
yields $gt_r\simeq 68.23$ for $\Delta/g=2$ and $gt_r\simeq 86.70$
for $\Delta/g=5$. Smaller revival times are possible for smaller values
of $\langle n\rangle$, but in general the observation of revivals will be
quite challenging, as they require $\Delta\sim g\gg \kappa_{\rm cav}$.

For large tunnel splitting, $\Delta/g\gg 1$,
$\Omega_{\mathrm{tun}}=\Delta+Ng^2/(2\Delta)+\mathcal{O}((g/\Delta)^3)$,
and Eq.~\eqref{rhoLLresonant} reduces to $\rho_{LL}\simeq
\sin^2(\Delta t/2)$, which is identical to the tunneling of a
particle without internal structure. Equation~\eqref{rhoeeresonant}
reduces to a Rabi oscillation $\rho_{ee}\simeq \cos^2(\sqrt{N}g
t/2)$.

\subsection{Non-resonant atom-field interaction}

\begin{figure}
\begin{center}
\includegraphics[width=.8\linewidth]{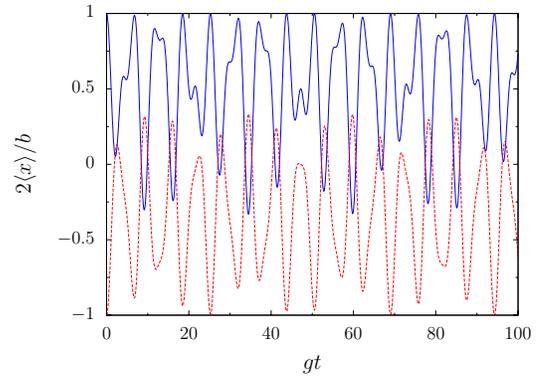}
\end{center}
\caption{(Color online) Average position of the atom in the double
well as a function of time for $\Delta=\delta=g$, $\kappa=\pi/4$ and
$N=1$. The blue solid/red dashed curve corresponds to an excited
atom initially located in the left/right well.} \label{rhoLL_qp}
\end{figure}

For non-resonant atom-field interaction ($\delta\ne 0$), and
intermediate tunnel splitting [see Fig.~\ref{rhoLL_qp} for
$\Delta=\delta=g$], $\langle x(t) \rangle$ involves in general the
two non-commensurate frequencies $\Omega_+$ and $\Omega_-$ and
varies therefore quasiperiodically as a function of time.
Figure~\ref{rhoLL_qp} also shows that an atom initially located in
one of the two wells remains mostly confined to that well.

\begin{figure}
\begin{center}
\includegraphics[width=.8\linewidth]{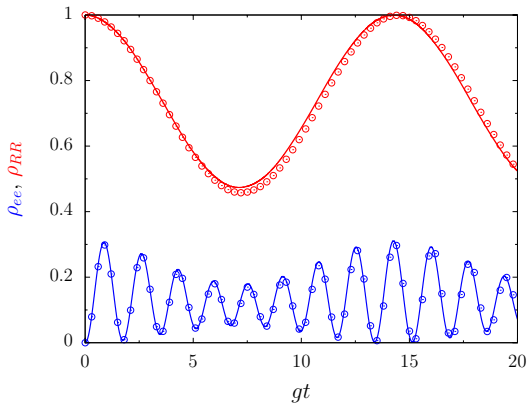}
\end{center}
\caption{(Color online) Density matrix elements $\rho_{RR}$ (top)
and $\rho_{ee}$ (bottom) as a function of the interaction time $gt$
for an initially excited atom located in the right well and for the
parameters $\Delta/g\simeq 0.3336$, $\delta/g=3$, $\kappa=\pi/4$,
and $N=1$. Numerical results from the propagation of the time
dependent Schr\"{o}dinger equation with Hamiltonian
(\ref{tHamiltonian}) and rotating wave approximation are represented
by circles and analytical results by solid curves. The time
propagation was done with ($\hbar=m=1$) $g=0.01$ and the double well
potential $V(x)=0.08x^4-x^2$ yielding a tunnel splitting
$\Delta\simeq 0.003336$ and a ratio
$\tilde{\Delta}/\sqrt{4g^2+\delta^2}\simeq 44.4\gg 1$. }
\label{rhoLLrhoee}
\end{figure}

For small tunnel splitting, $\Delta/g\ll 1$ and large detuning
$|\delta|/g\gg 1$ (with $\Delta|\delta|/g^2\sim 1$), the matrix
elements of $U$ simplify to
\begin{subequations}\label{twophoton}
\begin{align}
    U_{13} ={}& \frac{i\,Ng^2\sin
    2\kappa}{\sqrt{\delta^2\Delta^2+N^2g^4\sin^2(2\kappa)}}\:
    \sin\left(\frac{\bar{\Omega}t}{2}\right)\\
    U_{33} ={}& \cos\left(\frac{\bar{\Omega}t}{2}\right)
    +\frac{i\delta\Delta}{\sqrt{\delta^2\Delta^2+N^2g^4\sin^2(2\kappa)}}\sin\left(\frac{\bar{\Omega}t}{2}\right)
\end{align}
\end{subequations}
up to corrections of order $\mathcal{O}(\Delta/g)$ and a phase
factor $e^{i[(Ng^2/\delta+\delta)-(2N-1)\omega] t/2}$ while the
components $U_{12}$ and $U_{23}$ are of order
$\mathcal{O}(\Delta/g)$. In this situation, the system oscillates
only between the two states $|N-1,+,e\rangle$ and $|N-1,-,e\rangle$
with a single frequency
\begin{equation}
\bar{\Omega}=\frac{\sqrt{\delta^2\Delta^2+
    N^2g^4\sin^2(2\kappa)}}{\delta},
\end{equation}
just as a three-level atom undergoing a Raman transition in the far
detuned regime behaves as a two-level system.

If the system is initially in the state $|N-1,-,e\rangle$, we have
from Eqs.~\eqref{twophoton}
\begin{equation}\label{rhoLLtwophoton}
    \rho_{LL}=\frac{1}{2}-\frac{N\delta\Delta\sin (2\kappa)}{2 {\bar{\Omega}}^2(\delta/g)^2}
    \left[1-\cos\left(\bar{\Omega} t\right)\right],
\end{equation}
and $\rho_{ee}=1$. For a detuning $\delta=\pm
Ng^2\sin(2\kappa)/\Delta$,
\begin{equation}\label{rhoLLtwophotonN}
    \rho_{LL}=\frac{1}{2}\mp\frac{1}{4}\left[1-\cos\big(\sqrt{2}\Delta
    t\big)\right].
\end{equation}

This regime may be suitable for coherently manipulating the atom
position through access to its internal degrees of freedom with a
laser. Coherent manipulation of the position of neutral atoms has
been proposed and demonstrated before, see
e.g.~\cite{Mandel03,Mompart03,Sebby06}. In these examples, the
manipulation is done by modifying the external potential. The
mechanism we propose here is very different, as the potential
remains totally unchanged, and only internal transitions and the
tunneling effect are used to move the atom in a controlled way. As
an example, we show how the atom can be prepared in the left well
starting from the ground state $|0,-,g\rangle$ for
$\delta=-g^2/\Delta$.

 We first
apply a $\pi$-pulse with an external laser resonant with the atomic
transition. By using a laser with a wave vector perpendicular to the
$Ox$-direction, only the atomic internal degree of freedom is
affected, resulting in the transition $|0,-,g\rangle \to -i
|0,-,e\rangle$. We assumed that the laser Rabi frequency $\Omega_R$
is much larger than the tunnel frequency $\Delta$.

Now we use the coupling between the internal and external degrees of
freedom to create a superposition of the $|0,\pm,e\rangle$ states,
and then apply a second resonant $\pi$-pulse to get back to the
uncoupled states $|0,\pm,g\rangle$. For $\Delta/g\ll 1$,
$\delta=-g^2/\Delta$ and $\kappa=\pi/4$, the initial state
transforms according to
\begin{equation}
   |0,-,g\rangle \xrightarrow[\Omega_Rt=\pi]{} |0,-,e\rangle
   \xrightarrow[\Delta t=\pi/\sqrt{2}]{} |0,L,e\rangle
   \xrightarrow[\Omega_Rt=\pi]{} |0,L,g\rangle
\end{equation}
up to a physically irrelevant phase. Other coherent
superpositions of $|0,+,g\rangle$ and $|0,-,g\rangle$ can be obtained
by choosing appropriate interaction times.

In order to verify that the two-level approximation for the external
motion used in the derivation of the Hamiltonian is a good
approximation, we have numerically solved the time dependent
Schr\"{o}dinger equation with Hamiltonian (\ref{tHamiltonian}) and
rotating wave approximation but with the exact external potential
$V(x)$ (i.e.\ with a large number of vibrational states).
Figure~\ref{rhoLLrhoee} shows that provided $\tilde{\Delta}\gg
\sqrt{4g^2+\delta^2}$ as stated before, to take only the two lowest
vibrational states into account is indeed a good approximation.

We finally comment on possible experimental realizations of our
model. Double well potentials with tunable well-to-well separation
have been demonstrated with optical dipole traps e.g.~in
\cite{Shin04,Sebby06}, and on atom chips e.g.~in
\cite{Hinds01,Haensel01}. For our model, the double well potential
has to be realized inside the cavity. Optical trapping and even
cooling of atoms close to their ground state inside a cavity has
been achieved in several groups by now
\cite{Ye99,McKeever03,Sauer04,Maunz04}, but up to our knowledge
double well potentials have not been realized in a cavity so far.
However, some of the cavities developed have a very long lateral
opening (up to 222 $\mu$m \cite{Fortier07}) and should allow more
complicated trapping potentials (optical lattices intersecting a
cavity have been realized in Chapman's group \cite{Fortier07}). We
remark that it is not essential for our model that the double well
potential be aligned with the cavity axes. Any other orientation is
possible, and only leads to modified coefficients $\cos\chi
\sin\kappa$ and $\sin\chi \cos\kappa$.

At certain ``magical wavelengths'', Cs, Yb, Sr, Mg, and Ca atoms in
optical traps experience the same potential for ground and excited
internal states coupled by a dipole transition
\cite{McKeever03,Katori03,Brusch06,Barber06}. In a symmetric
potential $V(z)$ the tunneling frequency $\Delta$ is given in WKB
approximation by $\Delta\sim\omega_{\mathrm{osc}}\exp(-1/\hbar
\int_{-a}^a\sqrt{2m(E_0-V(z))}\,dz)$ where $E_0$ is the ground state
energy, $\omega_{\mathrm{osc}}$ the single well harmonic oscillation
frequency, and $z=\pm a$ are the corresponding classical turning
points delimiting the range of the barrier. The exponential factor
can approach unity for a barrier that is only slightly higher than
the ground state energy $E_0$, in which case cooling to temperatures
$k_B T<\hbar \Delta$ should be possible with state of the art
techniques \cite{McKeever03}. In \cite{McKeever03} a trap depth
$V_0/\hbar=47$ MHz was achieved inside a cavity with $1.2$ mW laser
power. In any case, the trap frequency and thus the tunneling
splitting are determined by the laser power and the focussing (or
the wavelength for optical lattices), and can therefore be
controlled independently of $\Gamma$, $\kappa_{\rm cav}$, such that
there should be no fundamental problem achieving $\Delta\gg
\Gamma,\kappa_{\rm cav}$. The detection of the tunneling motion
should be possible by optical imaging, i.e.\ diffusion of laser
light from another transition in the optical regime with smaller
wavelength than the well separation. Alternatively, one might
monitor the transmission through the cavity in the case that it
differs for the two locations of the wells \cite{Maunz05}. Another
possibility might be using the atomic spin as a position meter
\cite{Haycock00}.

\begin{acknowledgments} We thank Jacques Vigu\'e for an interesting
discussion and CALMIP (Toulouse) for the use of their computers.
This work was supported by the Agence National de la Recherche
(ANR), project INFOSYSQQ, and the EC IST-FET project EUROSQIP.
\end{acknowledgments}

\bibliography{../../mybibs_bt}

\end{document}